\documentclass[prd,aps,epsfig,floats,superscriptaddress,showpacs]{revtex4}  
\usepackage{graphics}  
\tighten  
\newcommand{\ti}{\tilde}

\newcommand{\lt}{\left}
\newcommand{\rt}{\right}
 
\newcommand{\la}{\langle}
\newcommand{\rg}{\rangle}

\newcommand{\nn}{\nonumber}
\newcommand{\ra}{\vec}
\newcommand{\f}{\frac}

\newcommand{\be}{\begin{equation}}  
\newcommand{\ee}{\end{equation}}
\newcommand{\bea}{\begin{eqnarray}}
\newcommand{\eea}{\end{eqnarray}}
\newcommand{\bean}{\begin{eqnarray*}}
\newcommand{\eean}{\end{eqnarray*}}

\def\lsim{\raise 0.4ex\hbox{$<$}\kern -0.8em\lower 0.62ex\hbox{$\sim$}}  
\def\gsim{\raise 0.4ex\hbox{$>$}\kern -0.7em\lower 0.62ex\hbox{$\sim$}}

\begin{document}  
%\draft  
%\twocolumn[\hsize\textwidth\columnwidth\hsize\csname@twocolumnfalse\endcsname  

\title{Gravitational force in weakly correlated particle distributions}
    
\author{Andrea  Gabrielli}    
\affiliation{ ``E. Fermi'' Center, Via Panisperna 89 A, Compendio del
Viminale, 00184 - Rome, Italy }
\author{Adolfo Paolo Masucci}    
\affiliation{ Dipartimento di Fisica,  
      Universit\`a di Roma ``La Sapienza'' -   
            P.le A. Moro 2, I-00185 Roma, Italy} 
\affiliation{ Dipartimento di Fisica,
      Universit\`a di ``Roma Tre'', Italy}  
\author{Francesco Sylos Labini}    
\affiliation{Laboratoire de Physique Th\'eorique,    
         Universit\'e de Paris XI, B\^atiment 211,    
  91403 Orsay, France}
  
\begin{abstract}    
\begin{center}    
{\large\bf Abstract}    
\end{center}     
We study the statistics of the gravitational (Newtonian) force in a
particular kind of weakly correlated distribution of point-like and
unitary mass particles generated by the so-called Gauss-Poisson point
process. In particular we extend to these distributions 
the analysis {\em \`a la
Chandrasekhar} introduced for purely Poisson processes. In this way we
can find the asymptotic behavior of the probability density function
of the force for large values of the field as a generalization of the
Holtzmark statistics.  The validity of the introduced approximations
is positively tested through a direct comparison with the analysis of
the statistics of the gravitational force in numerical simulations of
Gauss-Poisson processes.  Moreover the statistics of the force felt by
a particle due to only its first nearest neighbor is analytically a
numerical studied, resulting to be the dominant contribution to the
total force.
\end{abstract}    
%05.40,02.50,98.80.-k }    
\maketitle    
\date{today}    
%]    
\twocolumngrid    
     
\section{Introduction}    
    
The knowledge of the statistical properties of the gravitational field
in a given distribution of point-particles with the same (unitary)
mass provides an important information about the system in many
cosmological and astrophysical applications in which the particles are
treated as elementary objects.  In fact such an
information acquires a particular importance in the contexts of
stellar dynamics and of the cosmological $n-$body simulations to study
the formation of structures from initial mass density perturbations
\cite{bsl02,bjsl02}.  Similar studies are involved in other domains of
Physics, such as the analysis of the statistics of the {\em
dislocation-dislocation} interaction for what concerns the analysis of
crystal defects in condensed matter physics
\cite{disl}. Until now a complete study of this problem has been
accomplished only in the case of an uncorrelated Poisson particle
distribution
\cite{gnedenko,chandra}. Partial results have been found more recently in 
other two cases: 1) a fractal point distribution \cite{our}, and 2) a
radial density of particles \cite{delpo}.  In this paper we study the
case of a so-called Gauss-Poisson (GP) point process
\cite{stoyan,kerscher}, which generates particle distributions
characterized by only two-points correlations, i.e. connected
$n-$points correlation functions vanish for $n\ge 3$.  In this sense
it can be seen as the first step of correlated systems beyond the
completely uncorrelated Poisson distribution. For the GP point process
we generalize the method used by Chandrasekhar \cite{chandra} for the
Poisson case introducing some approximations.  Moreover we study the
contribution to the total force experienced by a particle due to the
first nearest neighbor (NN), in order to evaluate the weight of the
granular neighborhood of a fixed particle.  Indeed, as clearly
discussed by Chandrasekhar \cite{chandra}, one the main problem of the
dynamics of a self gravitating particle distribution is concerned with
the analysis of the force acting on a single particle.  Such a study
is at the basis of the analysis of particle and fluid dynamics.  In a
general way, it is possible to show that there are two different
contributions: the first is due to the system as a whole and the
second is due to the influence of the immediate neighborhood of the
particle.  The former is a smoothly varying function of position and
time while the latter is subject to relatively rapid
fluctuations. These fluctuations, which are then the subject of the
present paper, are related to the underlying statistical properties of
particle distribution.

\section{Gravitational force probability density in a Poisson
distribution} 
    
Firstly, let us recall Chandrasekhar's \cite{chandra} results for the
Poisson case. A Poisson distribution of point-particles with average
density $n$ in a volume $V$ is obtained by occupying randomly with a
particle of unitary mass each volume element $dV$ with probability $n
dV$ (with $n>0$ and $ndV\ll 1$) or leaving it empty with complementary
probability $1-ndV$ with no correlation between different volume
elements. Therefore the average number of particles in the volume $V$
is $N=nV$ with fluctuations from realization to realization of the
order of $\sqrt{nV}$ (the so-called Poissonian fluctuation), which
become negligible with respect to $N$ in the large $V$ limit. By
definition the connected two-point correlation function $\tilde
\xi(\vec{r})$ has only the diagonal part, that is $\tilde
\xi(\vec{r})=\delta(\vec{r})/n$.  Any other statistically homogeneous
particle distribution is characterized by a connected two-points
correlation function of the form
\cite{hansen,hz} 
\be
\tilde \xi(\vec{r})=\frac{\delta(\vec{r})}{n} +\xi(\vec{r})
\ee
where $\xi(\vec{r})$ is the non-diagonal part due to correlations
between the positions of different particles.

\begin{figure}[tbp]      
\scalebox{0.35}{\includegraphics{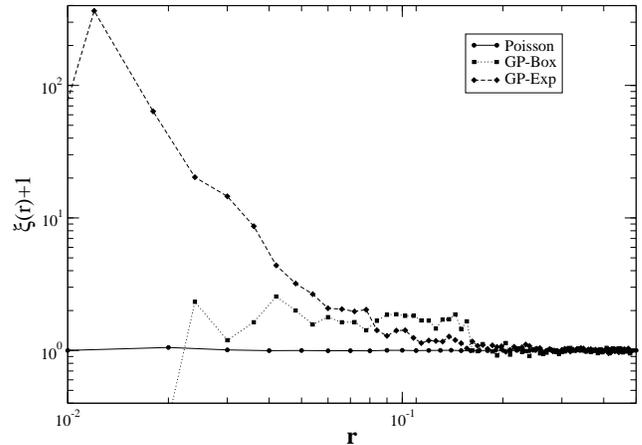} }     
\caption{Connected two-points correlation function plus one 
measured in different 
types of the particles distribution. The correlations function are taken by
averaging over $1000$ realizations of a sample of $N=10^2$ particles (in the
same fixed volume $V=1$.)
%{\it Top panel:} We show the  
Poisson case,  the
Gauss-Poisson case (GP-box) where  $p(r)$ is a $box-function$. 
and the 
Gauss-Poisson case (GP-Exp), where  
$p(r)=\f{1}{4\pi r_0}\frac{\exp (-\frac{r}{r_{0}})}{r^{2}}$. 
\label{fig1} }     
\end{figure}

In general, given the particle distribution,
the gravitational field acting on the origin of axis is given by:
\be
\vec{F}=\sum_i \frac{\vec{r}_i}{r_i^3}\,,
\label{eq1}
\ee
where the sum runs over all the system particles. Once the statistical
ensemble of particle distributions is chosen, one can evaluate the
probability density function (PDF) $\tilde W(\vec{F})$ of the field
$\vec{F}$ by taking the average of $\delta(\vec{F}-\sum_i
\frac{\vec{r}_i}{r_i^3})$ over the ensemble. In particular for a Poisson
distribution in a volume $V$ with average density of particles $n$
this calculation can be performed in the following way
\cite{chandra}. Since in this case the positions of different
particles are not correlated at all, the joint PDF of the positions of
the $N$ particles of the system is simply given by:
\be
p \left(\vec{r}_1,\vec{r}_2, ..., \vec{r}_N\right)=\prod_{i=1}^{N}\f{1}{V}
=\f{1}{V^N} \;.
\label{poisson-pos}
\ee
Therefore the PDF of the total gravitational force acting on the origin is: 
\[\tilde W(\vec{F})=
\int_V...\int_V \left[\prod_{i=1}^{N}\frac{d^3r_i}{V} \right]
\delta\left(\vec{F}-\sum_{i=1}^N \frac{\vec{r}_i}{r_i^3}\right) \;.\]
By using the Fourier representation of the Dirac delta function and taking 
the thermodynamic limit $V,N\rightarrow +\infty$ with $N/V=n$, we
obtain:
\be 
\label{eq2}
\tilde W(\vec{F})=\f{1}{(2\pi)^3}\int d^3k e^{i\vec{k}\cdot\vec{F}-
nC_P(k)}\,,
\ee
where 
\be 
\label{eq3}
C_P(k)=\frac{4}{15}(2\pi k)^{3\over 2}\,.
\ee
Note that $A(k)=\exp(-nC_P(k))$ is the Fourier transform
of $\tilde W(\vec{F})$, i.e. $A(k)$ is the so-called
{\em characteristic function} of the stochastic field $\vec{F}$ 
\cite{gnedenko}. Since $C_P(k)$ depends only on $k=|\vec{k}|$,
Eq.~\ref{eq2} says that the direction $(\theta,\phi)$ of $\vec{F}$ 
is completely random (i.e. with a PDF $p(\theta,\phi)=1/(4\pi)$)
and decoupled from $F=|\vec{F}|$, whose PDF (defined with $F\ge 0$) is 
instead given by
\be 
\label{eq4}
W(F)=\f{2 F}{\pi}\int_0^{+\infty}dk k\sin(kF)\
\exp\lt(-\f{4n}{15}(2\pi k)^\f{3}{2}\rt)\,.
\ee
This important result is known under the name of {\em Holtzmark
distribution}. An explicit expression of $W(F)$ is not possible to be
obtained; anyway it is rather simple to study the asymptotic regimes.
Probably the most important feature of Eq.~\ref{eq4} is the behavior
at large $F$ that can be found to be $W(F)\simeq 2\pi n F^{-\f{5}{2}}$
for $F\gg n^{-2/3}$.  Now we show that this limit behavior is
completely determined by the position of the (NN) particle.  In order
to show this result we have to evaluate the probability $\omega(r)dr$
that, given a particle, its first NN particle is at a distance between
$r$ and $r+dr$. An equation satisfied by $\omega(r)$ can be found by
considering that the probability of finding the first neighbor between
$r$ and $r+dr$ is equal to the product of the probability that there
is no particle in the distance interval $(0,r]$ and the probability
$4\pi n r^2 dr$ of finding a generic particle in the interval of
distances $(r,r+dr]$ \cite{hertz}, that is
\be \label{350}
\omega(r)=\lt(1-\int^r_0{\omega(x)dx}\rt)4\pi r^2n\,.
\ee
The derivation of Eq.~\ref{350} is based on the fact that there is no 
correlation between the position of different particles, implying that 
the probability of finding no particle in $(0,r]$ is independent of 
the probability of finding a particle in $(r,r+dr]$. This of course 
holds for a homogeneous Poisson distribution, but in general it is not 
true for correlated distributions.
Equation \ref{350} can be simply solved to give:
 \be \label{351}
\omega(r)=4\pi n r^2 \exp \left(-\f{4\pi}{3}nr^3\right)\,.
\ee
By considering that $F=1/r^2$, we can find, by a simple change of
variable, the PDF of the modulus of the gravitational field generated
by the first neighbor as:
\be \label{357}
W_{nn}(F)=2\pi F^{-\f{5}{2}}n\exp\lt(-\f{4\pi F^{-\f{3}{2}}n}{3}\rt)\,.
\ee
In the limit $F\gg n^{2/3}$ Eq.~\ref{357} reads
\be \label{349a}
W_{nn}(F)\simeq2\pi nF^{-\f{5}{2}}\;,
\ee
which is exactly the same of the asymptotic behavior of the PDF $W(F)$
of the modulus of the total force.  This result implies that in a Poisson
distribution (see Fig.\ref{fig1}), in which large-scale correlations are
absent but density fluctuations are present at all scales, 
the main contribution to the force acting on a particle comes
from particles in its neighborhood, the rest of faraway
particles giving only a finite additional contribution
because of statistical isotropy.

\section{The Gauss-Poisson point process}

%The aim of the present paper is to 
We now discuss the statistical properties
of the gravitational Newtonian field arising by an infinite weakly
correlated particle system, and in particular of a so-called
Gauss-Poisson distribution of point-like field sources (of unitary
mass).  A GP particle distribution \cite{stoyan,kerscher} is built in
the following way: first of all, let us take a Poisson distribution of
particles of average density $n_0>0$.  The next step is to choose randomly a
fraction $0<q\le 1$ of these Poisson points and to attach to each of
them a new ``daughter'' particle in the volume element $d^3r$ at vectorial
distance $\vec{r}$ from the ``parent'' particle with probability
$p(\vec{r})d^3r$ independently one each other.  Therefore the 
net effect of this algorithm is of substituting a fraction $q$ of
particles of the initial Poissonian system with an equal number of
correlated binary systems. This is the reason why this kind of point
distribution can be very useful in all the physical applications
characterized by the presence of binary systems.

It is simple 
to show that the final particle density in the so-generated
GP distribution is $n=n_0(1+q)$.
It is also possible to show that the connected two-point correlation 
function is
\be
\tilde \xi(\vec{r})=\f{\delta(\vec{r})}{n}+\f{2q}{n(1+q)}p(\vec{r})
\label{xi-gp}
\ee
and that all the other connected $n-$points correlation function with
$n\ge 3$ vanish \cite{daley}. This means that all the statistics of a
GP stochastic distribution is reduced to the knowledge of one and
two-points correlations. For this reason the GP particle distribution
is the discrete analogous of the continuous Gaussian continuous
stochastic fields.  Moreover since $p(\vec{r})$ is a PDF,
$\tilde \xi(\vec{r})$ is non-negative and integrable over all the space,
hence correlations are short ranged.  This is the reason why the GP
point-particle distributions can be seen as the most weakly correlated
particle system beyond the Poisson one.  To show the validity of
Eq.~\ref{xi-gp} is a quite simple task, in fact it is sufficient to use
the definition of average {\em conditional} density $n_p(\vec{r})$ of
particles seen by a generic particle of the system at a vectorial
distance $\vec{r}$ from it without counting the observing particle
itself. It is simple to show that \cite{hz}:
\be
n_p(\vec{r})=n[1+\xi(\vec{r})]\,,
\label{n_p}
\ee
where $\xi(\vec{r})$ is the non-diagonal part of $\tilde \xi(\vec{r})$.
On the other hand in the GP model the conditional average 
density can be evaluated in the
following way: the number of particles seen in average by the
chosen particle in the origin in the volume element $d^3r$ around
$\vec{r}$ is $n d^3r$ if the chosen particle is neither a ``parent''
nor a ``daughter'' (i.e. with probability $(1-q)/(1+q)$) and $n d^3r+
p(\vec{r})d^3r$ if it is either a ``parent'' or a ``daughter''
(i.e. with a complementary probability $2q/(1+q)$). This gives directly
\[n_p(\vec{r})=n\left[1+\f{2q}{n(1+q)}p(\vec{r})\right]\,,\]
which is equivalent to Eq.~\ref{xi-gp}.  Note that if $p(\vec{r})$
depends only on $r$ (i.e. it is spherically symmetric) then the
particle distribution, in addition to be statistically homogeneous
(i.e. translational invariant), is also statistically isotropic
(i.e. rotational invariant).

\section{Generalization of the Holtzmark distribution to the 
Gauss-Poisson case}

We can now try to generalize the Holtzmark distribution to this
correlated case. Let us suppose of having generated a GP distribution
with fixed $n>0$ and $0<q\le 1$ in a volume $V$ (therefore $N=nV$ with
negligible relative fluctuations in different realizations of the
distribution in the infinite volume limit). Let us also set the
coordinate system in such a way that the origin is occupied by a
particle of the system.  We want to calculate the PDF
$\tilde{W}(\vec{F})$ of the total gravitational field $\vec{F}$
acting on the origin of coordinates due to all the particles out of
the origin conditioned to the fact that this point is occupied by a
particle of the system
\footnote{Note that in the Poisson case we did not evaluated the conditional
PDF of the gravitational field, because we did not impose the
occupation of the origin of coordinates by a particle of the
system. However in the Poisson case there is no difference between
conditional and unconditional PDF because there is no correlation
between the position of different particles}.
Therefore if the particles seen by the one in the origin are $N-1$
and $p_c(\vec{r}_1, \vec{r}_2,...,\vec{r}_{N-1})$ is the PDF of their 
position with respect the origin we can write:
\bea
\tilde{W}(\vec{F})=\int_V...\int_V \left[\prod_{i=1}^{N-1} d^3r_i
\right]p_c(\vec{r}_1, \vec{r}_2,...,\vec{r}_{N-1}) \cdot 
\\ \nonumber 
\delta\left(\vec{F}-\sum_{i=1}^{N-1}\f{\vec{r}_i}{r_i^3}\right)
\label{gp-w}
\eea
The main problem to face is due to the fact that, since in the GP
case, two-points correlations are present, $p_c(\vec{r}_1,
\vec{r}_2,...,\vec{r}_{N-1})$ cannot be written as a product $N-1$ one
particle PDF's as in the Poisson case.  This property would prevent
from the possibility of applying the Markov method to this case, and
an explicit evaluation of $\tilde{W}$ would become impossible.  For this
reason we introduce the approximation consisting in imposing the
factorization
\[p_c(\vec{r}_1,\vec{r}_2,...,\vec{r}_{N-1})=\prod_{i=1}^N
\tau(\vec{r}_i)\,,\] 
but taking into account the fact that in average the
particle in the origin sees a density of particles in the point
$\vec{r}$ given by Eq.~\ref{n_p} with $\xi(\vec{r})=\f{2q}{n(1+q)}p(\vec{r})$, 
i.e.
\be
\tau(\vec{r})=\f{1+\f{2q}{n(1+q)}p(\vec{r})}{V+\f{2q}{n(1+q)}}\,.
\label{p_eff}
\ee
This approximation permits to use the Markov method to
find $\ti W(\vec{F})$ which, in the limit $V\rightarrow +\infty$,
can be shown to be given by 
\be 
\label{9}
\ti W(\vec{F})=\f{1}{(2\pi)^3}
\int d^3F\exp\lt(i\ra k\cdot\ra F-nC_{GP}(\vec{k})\rt)\,,
\ee
where 
\be
C_{GP}(\vec{k})=C_P(k)+\f{2q}{n(1+q)}\int d^3r\, p(\vec{r})
\left(1-e^{-i\f{\vec{k}\cdot\vec{r}}{r^3}}\right)\,.
\label{c_gp}
\ee
As shown below by a direct comparison with the results of numerical
simulations, this approximation is quite accurate at least in the
asymptotic regime $F\rightarrow +\infty$.  Note that the function
$A(\vec{k})=\exp\left(-nC_{GP}(\vec{k})\right)$ is nothing else the
{\em characteristic} function of the total force $\vec{F}$ acting on
the particle in the origin in the GP case.  As aforementioned, if the
PDF $p(\vec{r})$ depends only on $r=|\vec{r}|$, the particle
distribution is statistically isotropic. Consequently, as for the
Poisson case, $\ti W(\vec{F})$ also depends only on $F=|\vec{F}|$ and
$A(\vec{k})$ on $k=|\vec{k}|$.  That is the direction of $\vec{F}$ is
completely random while the PDF of $F$ is given by (by recalling
$p(\vec{r})$ with $p(r)$ to put in evidence the dependence only on
$r$)
\bea
\label{9bis}
&&W(F)\equiv 4\pi F^2 \ti W(\ra F)\\
&&=\f{2 F}{\pi}\int_0^\8dk k\sin(kF)
\exp\lt\{-\f{4(2\pi)^{\f{3}{2}} nk^\f{3}{2}}{15}\right.\nonumber\\
&&-\left.
\f{8\pi q}{1+q}\int_0^\8drr^2p(r)
\lt[1-\f{r^2}{k}\sin\lt(\f{k}{r^2}\rt)
\rt]\rt\}\;.\nn
\eea
We limit the rest of the discussion to this isotropic case.  As for
the Poisson distribution, it is not possible to find an explicit form of
$\ti W(F)$ (or equivalently of $W(F)$). 
However we can connect their large $F$ behavior to the
small $r$ behavior of $p(r)$ and to that of the Poisson case.  In
order to do this, it is important to use the general properties of the
Taylor expansion of the characteristic function $A(\vec{k})$
to the lowest order greater than zero. In particular in this isotropic
case we use that \cite{singular}, if $\ti W(\ra F)\sim F^{-\alpha}$
at large $F$ (note that $\alpha>3$ in any case as $\ti W(\ra F)$
is a normalizable PDF) then
\bea 
\label{char} 
&&A(\ra k)=\int d^3F
\exp\lt(-i\ra k \cdot \ra F\rt)\ti W(\ra F)=\nn\\ && \, =
\left\{
\begin{array}{ll}
1-\f{1}{6} \la F^2\rg k^2&\mbox{ if }\alpha>5\\
 1-ak^{\alpha-3}&\mbox{ if }3<\alpha\leq5\;,
\end{array}
\right.
\eea
where $a>0$ is a constant characterizing the singularity.
Note that $\alpha>5$ implies that $\left< F^2 \right>$ is finite, 
and that for the Poisson case $\alpha=9/2$, and correspondingly
$A(\vec{k})\simeq 1-\f{4n}{15}(2\pi k)^{\f{3}{2}}$.

Therefore our strategy is to find $\alpha$ by connecting the expansion
given in Eq.~\ref{char} to the form of $p(r)$ and in particular to its small
$r$ behavior. Let us suppose that $p(r)\sim r^\beta$ at small $r$ 
(in any case $\beta>-3$ as $p(r)$ is a PDF of a three-dimensional
stochastic variable).
It is quite simple to show that at small $k$ the integral
\[I(k;\beta)=\int_0^\8drr^2p(r)\lt[1-\f{r^2}{k}\sin\lt(\f{k}{r^2}\rt)
\rt]\]
behaves as follows:
\be
I(k;\beta)\simeq \left\{
\begin{array}{ll}
c_1 k^{\f{3+\beta}{2}} & \mbox{if }\beta<1\\
c_2 k^2 & \mbox{if } \beta\ge 1\,,
\end{array}
\right.
\label{asym}
\ee
where $c_1$ and $c_2$ are two positive constants depending on
the specific form of $p(r)$.
Consequently, by inserting this result in Eq.~\ref{9bis}, we can distinguish
three cases for what concerns the asymptotic behavior of $\ti W(\ra F)$:
\begin{itemize}

\item For $\beta>0$ the dominating part in $A(\ra k)$ at small $k$
is exactly the same as in the Poisson distribution with the same average
density $n$, i.e.
\be
A(\vec{k})\simeq 1-\f{4n}{15}(2\pi k)^{\f{3}{2}}\,,
\label{as1}
\ee
which implies $\ti W(\ra F)\sim F^{-\f{9}{2}}$ (or equivalently
$W(F)=4\pi F^2\ti W(\ra F)\sim F^{-\f{5}{2}}$)
at large $F$ with
the same amplitude of the pure Poisson case.

\item For $\beta=0$ we have again a substantially Poissonian
behavior but the coefficient of the non zero order term receives a 
contribution from two-points correlations:
\be
A(\vec{k})\simeq 1-8\pi n\left(\f{(2\pi)^{\f{1}{2}}}{15}
+\frac{c_1q}{n(1+q)}\right)k^{\f{3}{2}}\,,
\label{as2}
\ee
which implies again $\ti W(\ra F)\sim F^{-\f{9}{2}}$ at large $F$
but with a larger amplitude than in the Poisson case.

\item for $\beta<0$ the small $k$ behavior of $A(\ra k)$ is 
completely changed, being
\be
A(\vec{k})\simeq 1-\frac{8\pi c_1 q}{1+q}k^{\f{3+\beta}{2}}\,,
\label{as3}
\ee
which gives $\ti W(\ra F)\sim F^{-\f{9+\beta}{2}}$ (or equivalently
$W(F)=4\pi F^2\ti W(\ra F)\sim F^{-\f{5+\beta}{2}}$)
\end{itemize}

\section{Comparison with simulations} 

\begin{figure}[tbp]      
\scalebox{0.3}{\includegraphics{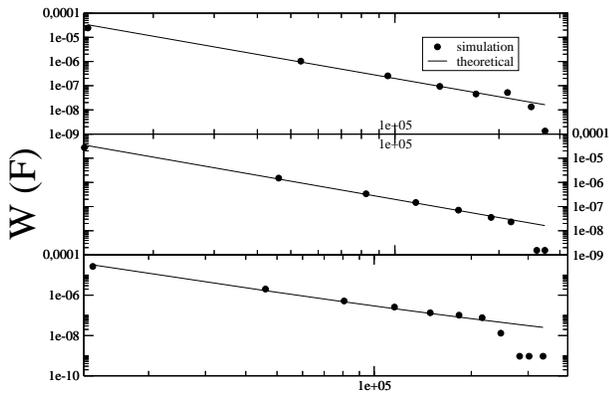}}     
\caption{Comparison between theoretical predictions 
(continuous lines) and simulations of the  
tail of the probability density function of the modulus of the
gravitational force $W(F)=4\pi\ti W(\vec {F})$. 
The average for the PDF is performed 
over $2\cdot 10^4$ realizations of a the same distribution with 
each of $N=10^5$  particles. 
{\it Top panel:} Poisson case. 
{\it Middle panel:} 
Gauss-Poisson case, where  $p(r)$ is a $box-function$. 
{\it Bottom panel: }
Gauss-Poisson case, where  
$p(r)$ has an  exponential cut-off: 
$p(r)=\f{1}{4\pi r_0}\frac{\exp (-\frac{r}{r_{0}})}{r^{2}}$
\label{fig2} }     
\end{figure}

In order to check the validity of these theoretical results, we have
performed numerical simulations consisting in generating two kinds of
GP distributions of particles with two explicite choices of $p(r)$
(see Fig.~\ref{fig1}), and in measuring directly the PDF of 
$\ti W(\ra F)$ in these cases:\\ (1) In the first one the choice of
$p(r)$ is simply a positive constant up to a fixed distance $r_0$ and
zero beyond this distance:
\be \label{v48}
p(r)=
\left\{
\begin{array}{ll}
\f{3}{4\pi r_0^3} & \mbox{if } 0<r\le r_0 \\
0  & \mbox{if } r< r_0\,.
\end{array}
\right. 
\ee
That is the probability of attaching a ``daughter'' particle at a
distance between $r$ and $r+dr$ from its ``parent'' is $3 r^2dr/r_0^3$
if $r\le r_0$ and zero for $r>r_0$. As shown above this choice of $p(r)$
should give $W(F)\sim F^{-\f{5}{2}}$ at large $F$
but with a larger amplitude 
than the pure Poisson case.\\
(2) In the second case $p(r)$ decays exponentially fast at large $r$ 
but increases as $r^{-2}$ at small $r$, i.e.
\be 
\label{v49}
p(r)=\f{1}{4\pi r_0}\frac{\exp (-\frac{r}{r_{0}})}{r^{2}}\;\;.
\ee
This choice of $p(r)$ should give $W(F)\sim F^{-\f{3}{2}}$ at
large $F$.  The results of these simulations for the large $F$
behavior of $\ti W(\ra F)$ have been compared to the theoretical
previsions showing a good agreement (see Fig.~\ref{fig2}). 
Consequently the validity of the approximation at the base of these 
calculations has been positively tested.
The found validity of the relations between the small scale behavior
of $\xi(\vec{r})$ and the large scale behavior of $W(F)$ for the GP case
suggests that this approach can be extended to more general cases of
correlated particle distributions.

\section{Nearest Neighbor approximation for the Gauss-Poisson case}

As for the Poisson case, we now analyze, for GP distributions in
the statistical isotropic case, the importance of the first 
%nearest
%neighbor 
NN contribution to the total force felt by the particle in the
origin. To this aim we use again only the information that the average
conditional density $n_p(\vec{r})$ (which depends only on $r$) seen by
the particle in the origin is given by Eq.~\ref{n_p}.  Following
exactly the same reasoning done for the Poisson case with
$n_p(\vec{r})$ replacing the simple $n$, we can write:
\be
\label{50}
\omega(r)=\lt(1-\int^r_0{\omega(x)dx}\rt)4n\pi r^2 (1+\xi(r))\,,
\ee
where again $\xi(r)=\frac{2q}{n(1+q)}p(r)$.
Equation \ref{50} can be solved to give
\bea 
&&\omega(r)=4n\pi r^2 (1+\xi(r))\exp\lt[-4n\pi \int_0^r s^2 (1+\xi(s))\rt]
\nn\\&&
=\omega_P(r)+4n\pi r^2\exp\lt[-4n\pi \int_0^r s^2 \xi(s)\rt]
\nn\\&& 
+4n\pi r^2 \xi(r)\exp\lt[-4n\pi \int_0^r s^2 (1+\xi(s))\rt]\,,
\label{51}
\eea
where we have called $\omega_P(r)$ the function $\omega(r)$ found
above for the Poisson case.  By imposing $p(r)\sim r^{\beta}$ at small
$r$ and using again $F=1/r^2$ in order to pass from $\omega(r)$ to
$W_{nn}(F)$, it is simple to see that $W_{nn}(F)$ has the same
aforementioned scaling behavior at large $F$ of $W(F)$ for all the
permitted values of $\beta$ and with the same coefficient.  Therefore
also in the GP case we have that the main contribution to the force
felt by a particle in the system is due to its first NN.
This was somehow expected because the main change introduced by passing
from Poisson to GP distribution concerns mainly the introduction of
additional fluctuations of density in the neighborhood of any particle.

\section{Discussion} 

In conclusion we have studied the gravitational force distribution in
the so called Gauss-Poisson particle distributions, which can be
considered in some sense the most weakly correlated case beyond the
Poisson one. For this particle systems we have seen how to generalize
the methods developed for the Poisson case in order to find the PDF of
the gravitational force. The main result is that, in the GP case,
significative deviations from the Poisson behavior can be caused only
by the small scale behavior of two-points correlations, which can
introduce strong modification in the large force regime when diverging
at small distances.  This is confirmed by direct results in numerical
simulations.  Moreover, as in the Poisson case, we found that the main
contribution to the force felt by a generic particle is due mainly to
its neighborhood.  The importance of this work is twofold. (i)
firstly, this is the first case of statistically homogeneous
correlated particle distribution in which a systematic study of the
gravitational force {\em \`a la Chandrasekhar} is done, (ii) this
study suggest some basic ingredients to be used in future attempts of
extending the analysis to more complex correlated particle
distributions.

\section {Acknowledgments}    
We thank L. Pietronero and M. Joyce for useful
comments and discussions. 
A.G. acknowledges the Physics Department of the University ``La
Sapienza'' of Rome (Italy) for supporting this research. F.S.L.
acknowledges the support of EC Marie-Curie fellowship 
HPMF-CT-2001-01443

%\newpage     

\end{document}